\documentclass[RNAAS]{aastex63}

\submitjournal{RNAAS}

\shorttitle{DESI ELG TS}
\shortauthors{Raichoor et al.}

\usepackage{amsmath}	

\newcommand{\niceurl}[1]{\href{#1}{#1}}
\newcommand{\oii}{[\ion{O}{2}]} 
\newcommand{\zphot}{z_{\rm phot}} 

\begin{document}

\title{Preliminary Target Selection for the DESI Emission Line Galaxy (ELG) Sample}

\author[0000-0001-5999-7923]{Anand Raichoor}
\affiliation{Institute of Physics, Laboratory of Astrophysics, Ecole Polytechnique F\'{e}d\'{e}rale de Lausanne (EPFL), Observatoire de Sauverny, 1290 Versoix, Switzerland}
\author{Daniel J. Eisenstein}
\affiliation{Harvard-Smithsonian Center for Astrophysics, 60 Garden St., Cambridge, MA 02138}
\author[0000-0002-5652-8870]{Tanveer Karim}
\affiliation{Harvard-Smithsonian Center for Astrophysics, 60 Garden St., Cambridge, MA 02138}
\author[0000-0001-8684-2222]{Jeffrey A. Newman}
\affiliation{University of Pittsburgh, 100 Allen Hall, 3941 O'Hara St., Pittsburgh, PA 15260, USA}
\author[0000-0002-2733-4559]{John Moustakas}
\affiliation{Department of Physics \& Astronomy, Siena College, 515 Loudon Road, Loudonville, NY 12211, USA} 

\author{David D. Brooks}
\affiliation{Department of Physics \& Astronomy, University College London, Gower Street, London, WC1E 6BT, UK}

\author{Kyle S. Dawson}
\affiliation{Department of Physics and Astronomy, The University of Utah, 115 South 1400 East, Salt Lake City, UT 84112, USA}
\author[0000-0002-4928-4003]{Arjun Dey}
\affiliation{NSF's National Optical-Infrared Astronomy Research Laboratory, 950 N. Cherry Avenue, Tucson, AZ 85719, USA}
\author[0000-0002-2611-0895]{Yutong Duan}
\affiliation{Physics Department, Boston University, Boston, MA 02215, MA}
\author[0000-0002-8281-8388]{Sarah Eftekharzadeh}
\affiliation{Department of Physics and Astronomy, The University of Utah, 115 South 1400 East, Salt Lake City, UT 84112, USA}
\author[0000-0001-9632-0815]{Enrique Gazta\~naga}
\affiliation{Institute of Space Sciences (ICE, CSIC), 08193 Barcelona, Spain}
\affiliation{
Institut d\'~Estudis Espacials de Catalunya (IEEC), 08034 Barcelona, Spain
}
\author{Robert Kehoe}
\affiliation{Department of Physics, Southern Methodist University, 3215 Daniel Avenue, Dallas, TX 75275, USA}
\author[0000-0003-1838-8528]{Martin Landriau}
\affiliation{Lawrence Berkeley National Laboratory, 1 Cyclotron Road, Berkeley, CA 94720, USA}
\author{Dustin Lang}
\affiliation{Perimeter Institute, Waterloo, ON~N2L~2Y5, Canada}
\affiliation{Department of Physics and Astronomy, University of Waterloo, 200 University Ave W, Waterloo, ON N2L 3G1, Canada}
\author{Jae H. Lee}
\affiliation{Department of Physics, Harvard University, Cambridge, MA 02138, USA}
\affiliation{Vertex Pharmaceuticals, 50 Northern Ave., Boston, MA 02210, USA}
\author[0000-0003-1887-1018]{Michael E. Levi}
\affiliation{Lawrence Berkeley National Laboratory, 1 Cyclotron Road, Berkeley, CA 94720, USA}
\author[0000-0002-1125-7384]{Aaron M. Meisner}
\affiliation{NSF's National Optical-Infrared Astronomy Research Laboratory, 950 N. Cherry Avenue, Tucson, AZ 85719, USA}
\author{Adam D. Myers}
\affiliation{Department of Physics \& Astronomy, University of Wyoming, 1000 E. University, Dept 3905, Laramie, WY 8207}
\author[0000-0003-3188-784X]{Nathalie Palanque-Delabrouille}
\affiliation{IRFU, CEA, Universit\'e Paris-Saclay, F-91191 Gif-sur-Yvette, France}
\author{Claire Poppett}
\affiliation{Space Sciences Laboratory at University of California, 7 Gauss Way, Berkeley, CA 94720}
\author{Francisco Prada}
\affiliation{Instituto de Astrofisica de Andaluc\'{i}a, Glorieta de la Astronom\'{i}a, s/n, E-18008 Granada, Spain}
\author{Ashley J. Ross}
\affiliation{Center for Cosmology and AstroParticle Physics, The Ohio State University, Columbus, OH 43212}
\author[0000-0002-5042-5088]{David J. Schlegel}
\affiliation{Lawrence Berkeley National Laboratory, 1 Cyclotron Road, Berkeley, CA 94720, USA}
\author{Michael Schubnell}
\affiliation{Department of Physics, University of Michigan, 450 Church St., Ann Arbor, MI 48109, USA}
\author{Ryan Staten}
\affiliation{Department of Physics, Southern Methodist University, 3215 Daniel Avenue, Dallas, TX 75275, USA}
\author{Gregory Tarl\'e}
\affiliation{Department of Physics, University of Michigan, 450 Church St., Ann Arbor, MI 48109, USA}
\author[0000-0001-5191-2286]{Rita Tojeiro}
\affiliation{School of Physics and Astronomy, University of St Andrews, North Haugh, St Andrews, KY16 9SS, UK}
\author{Christophe Y\`eche}
\affiliation{IRFU, CEA, Universit\'e Paris-Saclay, F-91191 Gif-sur-Yvette, France}
\author[0000-0001-5381-4372]{Rongpu Zhou}
\affiliation{Lawrence Berkeley National Laboratory, 1 Cyclotron Road, Berkeley, CA 94720, USA}
\affiliation{University of Pittsburgh, 100 Allen Hall, 3941 O'Hara St., Pittsburgh, PA 15260, USA}

\begin{abstract}
DESI will precisely constrain cosmic expansion and the growth of structure by collecting $\sim$35 million redshifts across $\sim$80\% of cosmic history and one third of the sky to study Baryon Acoustic Oscillations (BAO) and Redshift Space Distortions (RSD).
We present a preliminary target selection for an Emission Line Galaxy (ELG) sample, which will comprise about half of all DESI tracers.
The selection consists of a $g$-band magnitude cut and a $(g-r)$ vs.\ $(r-z)$ color box, which we validate using HSC/PDR2 photometric redshifts and DEEP2 spectroscopy. The ELG target density should be $\sim$2400 deg$^{-2}$, with $\sim$65\% of ELG redshifts reliably within a redshift range of $0.6<z<1.6$.
ELG targeting for DESI will be finalized during a `Survey Validation' phase.

\end{abstract}


\section*{Introduction}

DESI \citep{desi-collaboration16a} will measure spectroscopic redshifts for $\sim$35 million galaxies and quasars over $\sim$80\% of cosmic history and one third of the sky.
DESI will target multiple extragalactic tracers optimized for different redshift ranges, supplemented by significant stellar samples for calibration and Galactic science.
Relaxed selections will be tested during a preliminary `Survey Validation' phase to validate and optimize targeting for the DESI `main' survey.

This note outlines preliminary targeting for an Emission Line Galaxy (ELG) sample in the redshift range $0.6<z<1.6$, which will constitute approximately half of DESI extragalactic tracers.
DESI exploits the abundance of star forming galaxies at $z\sim1$--2 to target ELG tracers with a spectroscopic redshift that can be measured reasonably quickly. The high star formation rate at these redshifts produces identifiable emission lines without the need to detect a strong continuum.
A key spectroscopic diagnostic 
is the \oii~doublet  at $\lambda \lambda$ 3726,3729: the DESI spectrographs are designed to resolve this feature over the targeted ELG redshift range.
ELGs, which have been used as tracers in previous surveys (e.g., WiggleZ and eBOSS), will also underpin future BAO surveys, such as PFS \citep{takada14} and \textit{Euclid} \citep{laureijs11}.\\

\section*{ELG Target Selection}
The target selection will use $grz$~imaging from the Legacy Surveys \citep{dey19}.
Results presented here are based on Data Release 8 of the Legacy Surveys\footnote{\niceurl{http://legacysurvey.org/dr8/}}.
The DESI footprint is split into two regions; `North' (Galactic $b>0^\circ$ and Dec.\ $>32.375^\circ$) and `South'; the `North’ has a slightly different photometric system and is shallower (0.5 magnitudes) in the $g$- and $r$-bands. We therefore define slightly different cuts for the `North' and `South'.\\

First, we require a minimum photometric quality by enforcing at least one observation and a positive signal-to-noise ratio in each of $g$-, $r$-, and $z$-band.
We also require that targets are not in corrupted pixels, nor near bright or medium-bright stars, globular clusters, or large galaxies (\texttt{MASKBITS} is not set for bits 1, 5, 6, 7, 11, 12 or 13).

\begin{figure*}
\gridline{\fig{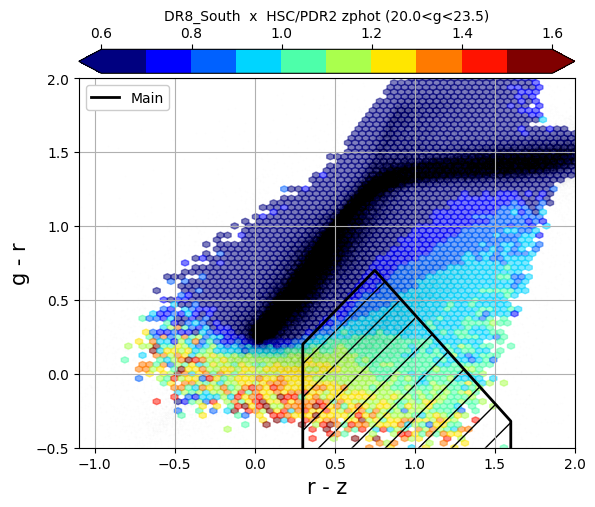}{0.5\textwidth}{(a)}
          \fig{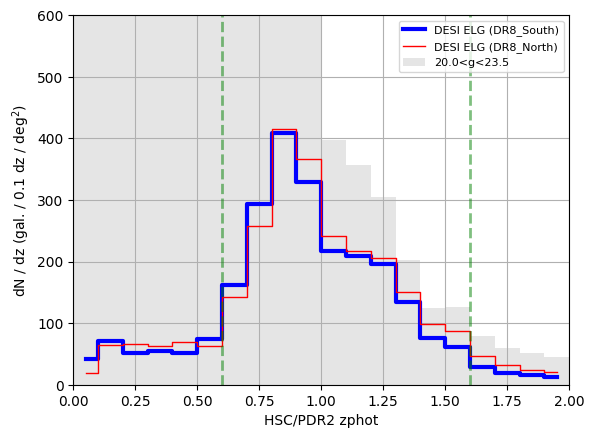}{0.5\textwidth}{(b)}
          }
\caption{
(a): $(g-r)$ vs.\ $(r-z)$ color-color diagram: the DESI ELG cuts for the `South' are displayed using black hatched lines. The colored hexagons represent the mean redshift for $20.0<g<23.5$ sources using a photometric redshift ($\zphot$) from the HSC/PDR2; photometric stars are displayed as black dots.
(b): HSC/PDR2 $\zphot$ distribution for the DESI ELG targets, in the `North' (red) and `South' (blue). The gray shaded histogram shows the $\zphot$ distribution of the parent $20.0<g<23.5$ sample. Dashed green lines show the desired redshift range.
\label{fig:1}}
\end{figure*}

\noindent Next, we apply the following cuts in $grz$ (see Figure~\ref{fig:1}):
\begin{subequations}
\begin{align}
20.0 < g < g_{\rm max} \label{eq:1a}\\
0.3 < (r - z) <  1.6 \label{eq:1b}\\
(g - r) < 1.15 \times (r - z) + \rm{zpt} \label{eq:1c}\\
(g - r) < -1.20 \times (r - z) +1.6 \label{eq:1d},
\end{align}
\end{subequations}
with ($g_{\rm max}$, zpt) = (23.6, -0.35) for the `North' and ($g_{\rm max}$, zpt) = (23.5, -0.15) for the `South'.
All magnitudes are corrected for Galactic extinction using the \citet{schlegel98} maps.
Eqs.\,(\ref{eq:1b}) and~(\ref{eq:1c}) select targets in the desired redshift range and Eq.\,(\ref{eq:1d}) favors star-forming galaxies.
As the photometry is noisier in the `North', our selection box is farther from the low-redshift locus to avoid significant contamination from $z<0.6$ galaxies.
Eq.\,(\ref{eq:1a}) targets the requisite \oii~flux \citep[see e.g.,][]{comparat15a} and also sets the density to $\sim$2400 deg$^{-2}$. 

As no spectroscopic truth table exists for a complete sample with $g \lesssim 23.5$, we assess our selection using HSC/PDR2 \texttt{DEmP} photometric redshifts \citep[$\zphot$;][]{aihara19} for the redshift distribution, and DEEP2 spectroscopic data over $0.8<z<1.4$ for the \oii~flux \citep{newman13a}.
The HSC/PDR2 $\zphot$ cover $\sim$100 deg$^2$ in the `North' and $\sim$200 deg$^2$ in the `South'. The $\zphot$ are estimated from deep $grizy$-photometry and are of exquisite quality for $z<1.6$ ELGs when compared to spectroscopy from eBOSS and from DESI Pilot Observations with the MMT \citep{raichoor20,karim20}.
Figure~\ref{fig:1} shows the $\zphot$ distribution of our ELG targets, demonstrating that $\zphot>1.0$ objects are efficiently selected; overall, $\sim$80\% of the selection has $0.6<\zphot<1.6$ for both `North' and `South'.

Finally, we characterize \oii~flux for our selection using measurements from \citet{comparat15a} in DEEP2 over $0.8<z<1.4$, where DEEP2 is complete over all fields for our $g < 23.5$--23.6 ELG sample.
We find that 76\% (`North') and 83\% (`South') of our targets have sufficient \oii~flux for a secure spectroscopic redshift measurement given the expected DESI specifications.\\

\section*{Conclusion}
This note outlines a preliminary DESI ELG selection based on a $g$-band cut and a $(g-r)$ vs. $(r-z)$ color-color box that produces $\sim$2400 deg$^{-2}$ targets.
Analyses using HSC/PDR2 $\zphot$ and DEEP2 \oii~flux show that $\sim$65\% of resulting ELGs will provide a reliable spectroscopic redshift within $0.6<z<1.6$, in accord with \citet{desi-collaboration16a}.
Preliminary ELG targeting will be tested during DESI `Survey Validation' to inform a final selection for the DESI `main' survey.
Target catalogs that use the selection described in this note are public\footnote{Available at \niceurl{https://data.desi.lbl.gov/public/ets/target/catalogs/} and detailed at \niceurl{https://desidatamodel.readthedocs.io}}.\\

AR acknowledges support from the ERC advanced grant LIDA and from the SNF grant 200020\_175751.
This research is supported by the Director, Office of Science, Office of High Energy Physics of the U.S. Department of Energy under Contract No.DE–AC02–05CH1123, and by the National Energy Research Scientific Computing Center, a DOE Office of Science User Facility under the same contract; additional support for DESI is provided by the U.S. National Science Foundation, Division of Astronomical Sciences under Contract No. AST-0950945 to the NSF’s National Optical-Infrared Astronomy Research Laboratory; the Science and Technologies Facilities Council of the United Kingdom; the Gordon and Betty Moore Foundation; the Heising-Simons Foundation; the French Alternative Energies and Atomic Energy Commission (CEA); the National Council of Science and Technology of Mexico; the Ministry of Economy of Spain, and by the DESI Member Institutions. The authors are honored to be permitted to conduct astronomical research on Iolkam Du’ag (Kitt Peak), a mountain with particular significance to the Tohono O’odham Nation.

\bibliography{ms}
\bibliographystyle{aasjournal}

\end{document}